\documentclass[aps,pra,reprint,showpacs,superscriptaddress]{revtex4-1}
\usepackage{graphicx}
\usepackage{dcolumn}
\usepackage{bm}
\usepackage{amssymb}
\usepackage{amsmath}
\usepackage{subfigure}
\usepackage{epstopdf}
\usepackage{float}
\usepackage{mathrsfs}
\usepackage{xcolor}
\begin{document}
\title{Control of Ultracold Atoms with a Chiral Ferromagnetic Film}
\author{Ren Qin}
\affiliation{School of Physics, Nankai University, Tianjin 300071, China}
\author{Yong Wang}
\email[]{yongwang@nankai.edu.cn}
\affiliation{School of Physics, Nankai University, Tianjin 300071, China}

\begin{abstract}
We show that the magnetic field produced by a chiral ferromagnetic film can be applied to control ultracold atoms. The film will act as a magnetic mirror or a reflection grating for ultracold atoms when it is in the helical phase or the skyrmion crystal phase respectively. By applying a bias magnetic field and a time-dependent magnetic field, one-dimensional or two-dimensional magnetic lattices including honeycomb, Kagome, triangular types can be created to trap the ultracold atoms. We have also discussed the trapping height, potential barrier, trapping frequency, and Majorana loss rate for each lattice. Our results suggest that the chiral ferromagnetic film can be a platform to develop artificial quantum systems with ultracold atoms based on modern spintronics technologies.   
\end{abstract}
\maketitle

\section{Introduction}
In the past three decades, the ultrocold atom physics has been greatly developed with the progressive techniques to cool, trap and manipulate neutral atoms with electromagnetic fields\cite{RMP1999,PhysUsp}. The quantum nature of the ultracold atom gases will emerge after suppressing the thermal fluctuations, and these artificially controllable quantum systems provide the ideal platforms for realizing Bose-Einstein condensation, simulating quantum many-body phenomena, performing quantum computation, designing atomic interferometers, \emph{etc}. Two typical mechanisms to control the neutral atoms are the alternating-current Stark energy shift in the high-frequency optical field and the Zeeman energy shift in the inhomogeneous magnetic field. The former has been widely applied to construct optical lattices for ultracold atoms, which play the crucial role in design various quantum simulators\cite{OptLatt}; while the latter lays the foundation to design and develop atom chips\cite{AtomChip}. In fact, the two mechanisms are more often combined together to achieve the best level of control of the ultracold atoms.   

Several strategies have been developed to produce the desirable magnetic field to control ultracold atoms. The most flexible approach is to fabricate current-carrying conductive microstructures to generate the Oersted field, which has been applied to guide and trap ultracold atoms in various configurations\cite{RMP2007}. This approach however suffers from the Johnson thermal noise and wire roughness for miniature devices. Another alternative way is to utilize the permanent ferromagnetic film with artifical patterns as the effective ``magnetization current"\cite{PRA2007,PRA2017}, which however is hard to be reconfigured and switched off. Besides, there also exist other theoretical proposals to generate magnetic lattice for ultracold atoms, such as the vortex array in the superconducting film\cite{NJP2010,PRL2013} and time-periodic magnetic field pulses\cite{NJP2015,PRL2016}.  

In addition to the artificial patterns, noncollinear magnetic textures can also induce inhomogeneous configurations of magnetic field  near the surface of the ferromagnetic film. Indeed, West \emph{et al.} have demonstrated the ability to manipulate ultracold $^{87}$Rb atoms by magnetic domain walls in planar magnetic nanowires\cite{NL2012}. Recently, the chiral ferromagnetic materials with finite Dzyaloshinskii-Moriya interaction(DMI) resulting from the lack spatial inversion symmetry\cite{Dzya,Moriya} have attracted broad research interests\cite{NRM1,NRM2}, mainly due to the discovery of magnetic skyrmions therein\cite{PRL2001,Nature2006,Science2009,Nature2010,NatMat2010,NatPhys2011} and the promising applications to develop spintronics devices\cite{NatNano2013,Proc2016}. With the formation of magnetic skyrmions, the magnetic field distribution near the surface of chiral ferromagnetic film will also been modified. For example, the magnetic field produced by an isolated magnetic skyrmion has been measured with the nitrogen-vacancy center in diamond\cite{NC2018}. The magnetic field distributions generated by the magnetic skyrmion crystals (SkXs) with different helicity have also been investigated theoretically\cite{NJP2018}. Therefore, it will be interesting to examine the possibility to control the ultracold atoms with the magnetic field produced by the chiral ferromagnetic film. 

In this paper, we show that the chiral ferromagnetic film can be utilized to design magnetic mirror, reflection grating, magnetic lattices, \emph{etc.}, which are the elementary devices to control the ultracold atoms. Considering the diversity of chiral ferromagnetic materials and the rapidly developing spintronics techniques, the approach proposed here could be crucial to develop ultracold atom physics in future.

\section{Basic Principle and Model}

\begin{figure}
\includegraphics[scale=0.5]{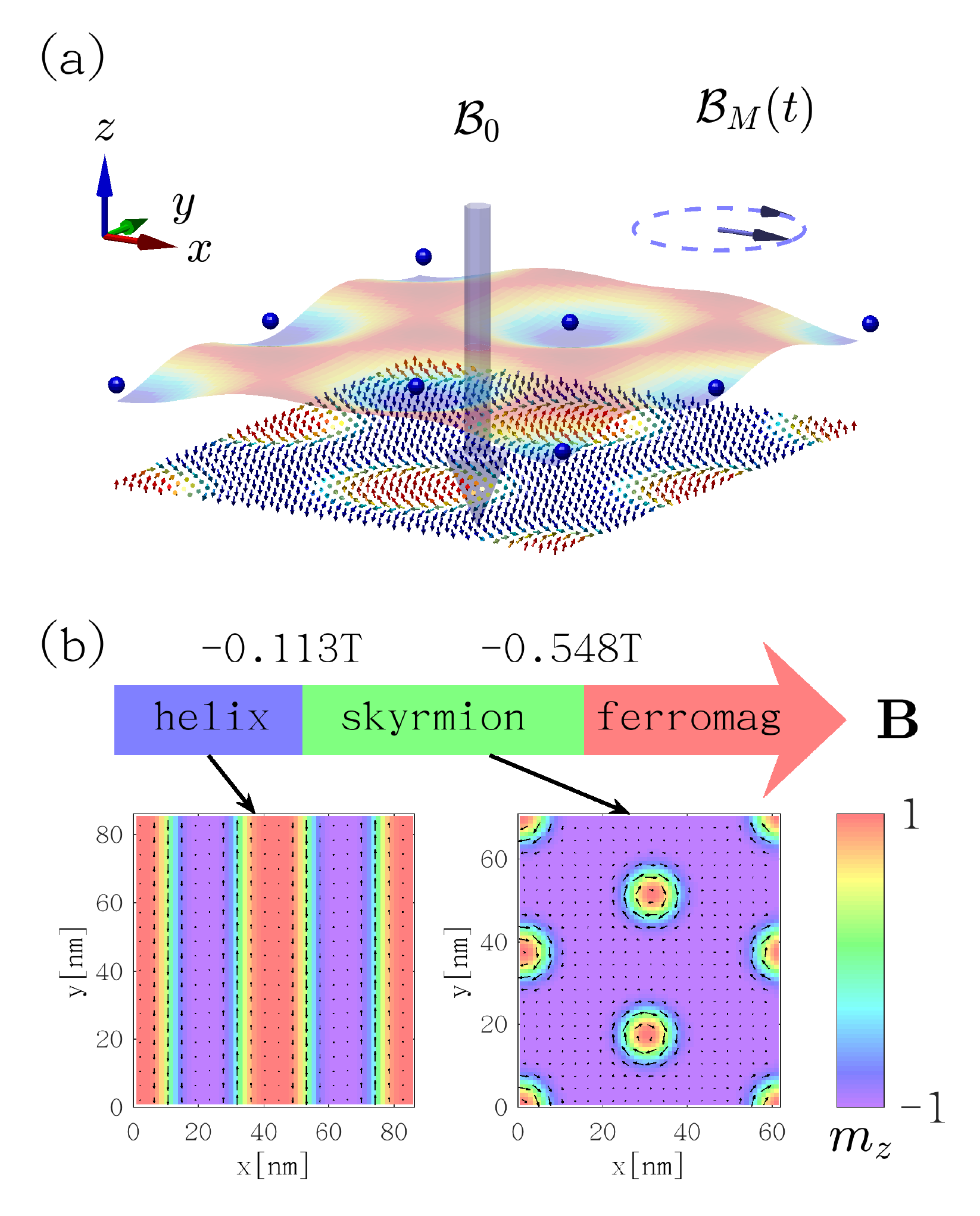}
\caption{(Color online) (a) Schematic diagram for the device. The chiral ferromagnetic film with prepared magnetization configuration is placed at the $x$-$y$ plane with $z=0$~nm, which will produce a magnetic field $\bm{\mathcal{B}}_{c}$; a bias magnetic field $\bm{\mathcal{B}}_{0}$ is applied perpendicular to the film; a time-dependent magnetic field $\bm{\mathcal{B}}_{M}(t)$ is applied parallel to the film. The trapping potential for the ultracold atoms (blue balls) is a combination of the three magnetic fields. (b) Phase transition of the chiral ferromagnetic film defined by Eq.~(\ref{Ham}). The magnetization configurations of helical (H) phase and skyrmion crystal (SkX) phase are obtained at $\mathbf{B}=0$~T and $\mathbf{B}=-(0,0,0.3)$~T respectively. The arrows denote the magnetization direction at each site, and the amplitude of $m_{z}$ is shown with different color.}\label{Fig1}
\end{figure}

The proposed device is schematically shown in Fig.~\ref{Fig1}(a). A chiral ferromagnetic film is placed in the $x$-$y$ plane with $z=0$~nm, which will generate the spatial distributions of magnetic field $\bm{\mathcal{B}}_{c}$. An uniform magnetic field $\bm{\mathcal{B}}_{0}$ is applied perpendicular to the film to create zero points at finite heights. Furthermore, a time-dependent magnetic field $\bm{\mathcal{B}}_{M}=\mathcal{B}_{M}(\sin\omega_{M}t,\cos\omega_{M}t,0)$ parallel to the film will be introduced to generate an effective time-orbiting potential in order to suppress the Majorana loss\cite{PRL1995,PRL2013}. Thus, the total magnetic field profile of such a device will be $\bm{\mathcal{B}}=\bm{\mathcal{B}}_{c}+\bm{\mathcal{B}}_{0}+\bm{\mathcal{B}}_{M}$. 

When an atom with magnetic dipole moment $\bm{\mu}$ is moving in the inhomogeneous magnetic field $\bm{\mathcal{B}}(\mathbf{r})$, it will experience the Stern-Gerlach force given by the potential energy $U(\mathbf{r})=-\bm{\mu}\cdot\bm{\mathcal{B}}(\mathbf{r})$. If the spatial change of the magnetic field felt by the atom is much slower than its Larmor precession, its magnetic moment will adiabatically follow the direction of the magnetic field, and the potential energy now depends only on the modulus of $\bm{\mathcal{B}(\mathbf{r})}$, which is given as $U(\mathbf{r})=m_{F}g_{F}\mu_{B}\mathcal{B}(\mathbf{r})$ for the hyperfine state $|F,m_{F}\rangle$\cite{AtomChip}. Here, $g_{F}$ is the Land\'{e} factor, $\mu_{B}$ is the Bohr magneton, $F$ is the total angular momentum quantum number, and $m_{F}$ is the magnetic quantum number of the atom. Therefore, the atom will be in the strong-field seeking state for $m_{F}g_{F}<0$ and in the weak-field seeking state for $m_{F}g_{F}>0$. Since no minimum can exist in the potential $U(\mathbf{r})$ for the strong-field seeking state according to the Earnshaw's theorem\cite{AtomChip}, only the atoms prepared in the weak-field seeking state can be magnetically trapped. For example, the $^{87}\text{Rb}$ and $^{7}\text{Li}$ atoms can be magnetically trapped when they stay at the state $|F=2,m_{F}=2\rangle$\cite{RMP2007,PRA2007-2}.

An external magnetic field $\mathbf{B}$ should be applied \emph{in advance} to tune the magnetization configuration of the chiral ferromagnetic film, which is described by the energy functional density  
\begin{eqnarray}
\mathcal{E}[\mathbf{m}]=\frac{J}{2}(\nabla\mathbf{m})^{2}+D\mathbf{m}\cdot(\nabla\times\mathbf{m})-K\mathbf{m}_{z}^{2}-M_{s}\mathbf{B}\cdot\mathbf{m}.\nonumber\\\label{Ham}
\end{eqnarray}
Here, $\mathbf{m}(\mathbf{r})$ is the magnetization distribution of the film normalized by the saturation magnetization $M_{s}$; the four terms in Eq.~ (\ref{Ham}) are the Heisenberg exchange interaction, the DM interaction, the perpendicular magnetic anisotropy energy, and the Zeeman energy in presence of the applied magnetic field $\mathbf{B}$ respectively. Then the magnetization dynamics will be given by the Landau-Lifshitz-Gilbert(LLG) equation
\begin{eqnarray}
\frac{d\mathbf{m}}{dt}=-\gamma\mathbf{m}\times\mathbf{H}^{eff}+\alpha\mathbf{m}\times\frac{d\mathbf{m}}{dt}.\label{LLG}
\end{eqnarray}
Here, $\gamma$ is the gyromagnetic ratio, $\mathbf{H}^{eff}=-\frac{1}{M_{s}}\nabla_{\mathbf{m}}\mathcal{E}[\mathbf{m}]$ is the effective magnetic field, and $\alpha$ is the Gilbert damping coefficient. For a given set of parameters, the stable magnetization configuration of the chiral ferromagnetic film is achieved as the stationary solution of the LLG equation\cite{NJP2018}. As shown in Fig.~\ref{Fig1}(b), when the external magnetic field $\mathbf{B}$ is perpendicular to the film and is continuously increased from zero, the ground state of the chiral ferromagnetic film will evolve from the helical phase to the skyrmion crystal phase at first, and reach the ferromagnetic phase finally. During the micromagnetic simulations, the parameters are set as $J=15$~pJ/m, $D=3$~mJ/m$^{2}$, $K=0.7$~MJ/m$^{3}$, $M_{s}=580$~kA/m, $\alpha=0.3$, and the thickness of the film is assumed as $1$~nm. The film is subdivided into cubic grids with the size $1\times1\times1$~nm$^3$, and the time step for the simulations is set as $0.01$~ps. Besides, the temperature is linearly decreased from $1000$~K to $0$~K during $100$~ns to avoid possible local minima trap. 

The external magnetic field $\mathbf{B}$ will be withdrawn once the desirable magnetization pattern is stably reached. Then the magnetic field $\bm{\mathcal{B}}_{c}(\mathbf{r})$ generated by the magnetization configuration $\mathbf{m}(\mathbf{r})$ of the chiral ferromagnetic film can be directly calculated as\cite{NJP2018}
\begin{eqnarray}
\bm{\mathcal{B}}_{c}(\mathbf{r})=\frac{\mu_{0}}{4\pi}\int d\mathbf{r}'\frac{3(\mathbf{m}(\mathbf{r}')\cdot\hat{\bm{\mathcal{R}}})\hat{\bm{\mathcal{R}}}-\mathbf{m}(\mathbf{r}')}{\mathcal{R}^{3}},\label{StrayField}
\end{eqnarray}
where $\bm{\mathcal{R}}=\mathbf{r}-\mathbf{r}'$. Below, we will study how to control the ultracold atoms in the weak-field seeking state with the magnetic field $\bm{\mathcal{B}}_{c}(\mathbf{r})$ when the chiral ferromagnetic film is prepared in the helical phase and skyrmion crystal phase respectively.

\section{Helical Phase}
\begin{figure}
\includegraphics[scale=0.5]{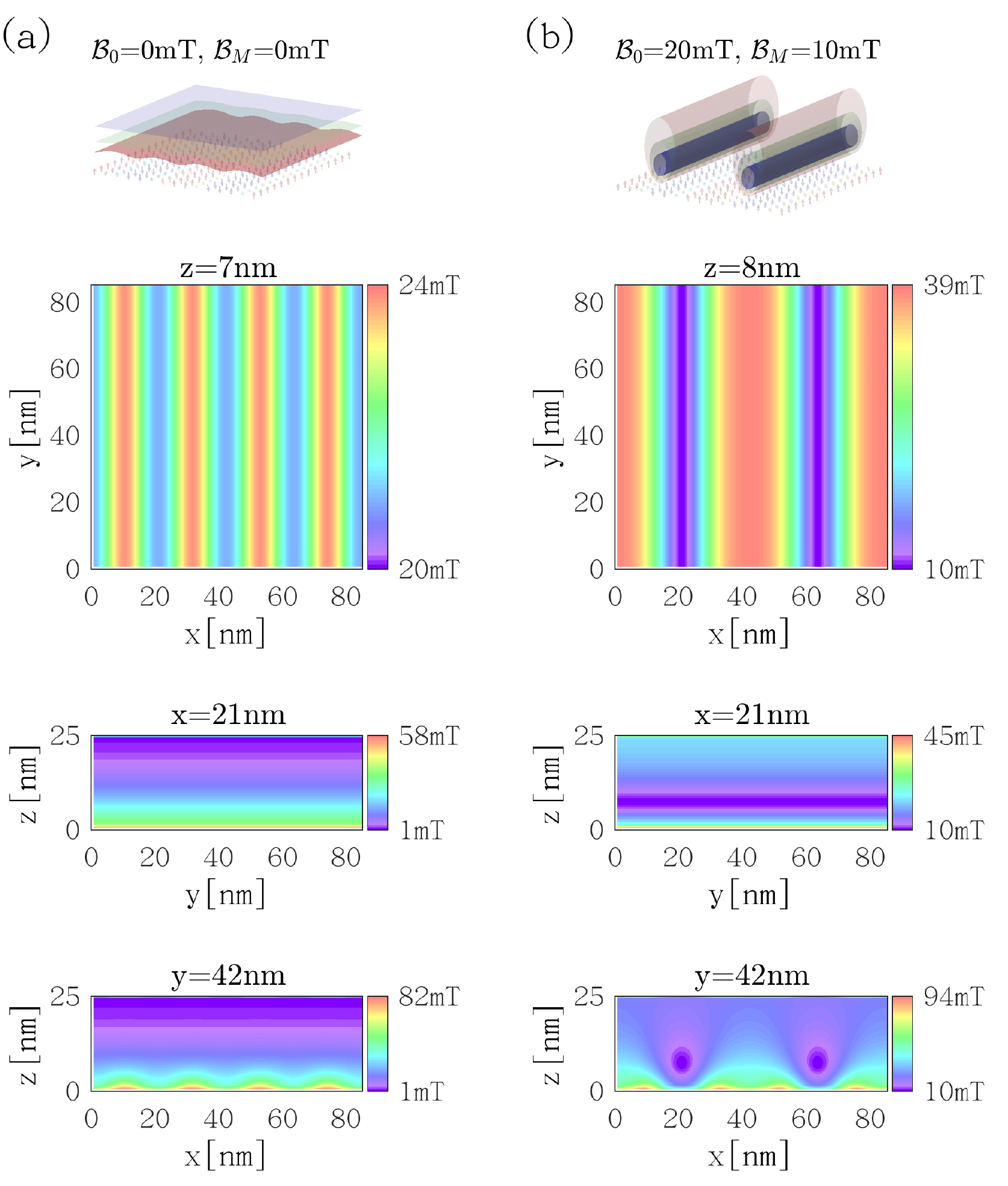}
\caption{(Color online) Profiles of magnetic field amplitude $\mathcal{B}(\mathbf{r})$ for the chiral ferromagnetic film at helical phase. (a) Magnetic mirror when $\mathcal{B}_{0}=0$~mT and $\mathcal{B}_{M}=0$~mT. Top : isosurfaces of $\mathcal{B}(\mathbf{r})$ with three values $8$~mT, $16$~mT, and $24$~mT; middle and bottom : cross sections of $\mathcal{B}(\mathbf{r})$ when $z=7$~nm, $x=21$~nm, and $y=42$~nm respectively. (b) One-dimensional magnetic lattice when $\mathcal{B}_{0}=20$~mT and $\mathcal{B}_{M}=10$~mT. Top : isosurfaces of $\mathcal{B}(\mathbf{r})$ with three values $13.5$~mT, $17.0$~mT, $20.5$~mT; middle and bottom : cross sections of $\mathcal{B}(\mathbf{r})$ when $z=8$~nm, $x=21$~nm, and $y=42$~nm respectively. }\label{Fig2}
\end{figure}

As shown in Fig.~\ref{Fig1}(b), the helical phase can be achieved as the ground state of the chiral ferromagnetic film when $\mathbf{B}=0$, where the magnetization configuration is periodically modulated in one direction and mimics the structure of a magnetic mirror for ultracold atoms fabricated from the permanent ferromagnetic materials\cite{AtomChip,JPD1999}. Nevertheless, it is advantageous to utilize the chiral ferromagnetic film in helical phase as a magnetic mirror, since the helical phase is spontaneously formed as the ground state and can also be easily switched off by applying an external magnetic field. Here, the complex fabrication processes are not required in comparison to the ferromagnetic film with artificial patterns\cite{AtomChip,JPD1999}. Besides, the period of the helical phase can be tuned by engineering the DMI\cite{NatNano2013-2,NatNano2015}, which thereby makes it flexible to control ultracold atoms. 

The magnetic field amplitude $\mathcal{B}_{c}(\mathbf{r})$ solely generated by the chiral ferromagnetic film in the helical phase in Fig.~\ref{Fig1}(b) has been numerically calculated based on Eq.~(\ref{StrayField}) and is demonstrated in Fig.~\ref{Fig2}(a). Here, the bias magnetic field $\bm{\mathcal{B}}_{0}$ and the time-dependent field $\bm{\mathcal{B}}_{M}$ have not been switched on yet. We see that the isosurfaces of $\mathcal{B}_{c}(\mathbf{r})$ will be nearly parallel to the $x$-$y$ plane. In fact, a weak periodic modulation of $\mathcal{B}_{c}(\mathbf{r})$ will happen at fixed height $z$. For example, $\mathcal{B}_{c}(\mathbf{r})$ will periodically vary from $21$~mT to $24$~mT when $z=7$~nm. On the other hand, $\mathcal{B}_{c}(\mathbf{r})$ will decay from $81$~mT to $2$~mT when the height $z$ increases from $0$~nm to $25$~nm. Therefore, the effective distance for the chiral ferromagnetic film to control the ultracold atoms will be tens of nanometers with the given simulation parameters here.

The simulated results in Fig.~\ref{Fig2}(a) can be further understood considering that the magnetization configuration of the helical phase can be approximated as a single-$\mathbf{Q}$ state
$\mathbf{m}_{h}(\mathbf{r})=\mathbf{m}_{0}\delta(z)+\mathcal{A}[\hat{\mathbf{e}}_{z}\cos(\mathbf{Q}\cdot\mathbf{r})+\hat{\mathbf{e}}\sin(\mathbf{Q}\cdot\mathbf{r})]\delta(z)$. Here, $\hat{\mathbf{e}}_{z}$ is the unit vector normal to the film and $\mathbf{Q}$ defines the wavevector of the helical state; the unit vector $\hat{\mathbf{e}}$ is determined as $\hat{\mathbf{e}}=\hat{\mathbf{e}}_{z}\times\hat{\mathbf{Q}}$ for the Bloch-type DMI given in Eq.~ (\ref{Ham}); $\mathcal{A}$ gives the amplitude of the modulated magnetization configuration of the helical state. The delta function $\delta(z)$ here implies that the thickness of the film is neglected. Then the magnetic field distribution $\bm{\mathcal{B}}_{h}(\mathbf{r})$ generated by the single-$\mathbf{Q}$ state $\mathbf{m}_{h}(\mathbf{r})$ is analytically obtained as\cite{NJP2018} 
\begin{eqnarray}
\bm{\mathcal{B}}_{h}(\mathbf{r})=\frac{\mathcal{A}Q}{2}e^{-Q|z|}[\text{sgn}(z)\hat{\mathbf{Q}}\sin(\mathbf{Q}\cdot\mathbf{r})+\hat{\mathbf{e}}_{z}\cos(\mathbf{Q}\cdot\mathbf{r})],\nonumber\\\label{mirror}
\end{eqnarray}
and its modulus is $\mathcal{B}_{h}(\mathbf{r})=\frac{\mathcal{A}Q}{2}e^{-Q|z|}$. Therefore, the chiral ferromagnetic film described by the single-$\mathbf{Q}$ state will establish an exponential repulsive potential to reflect the ultracold atoms at weak-field seeking state. The decay length of the potential is proportional to the period of the helical phase, which is usually about a few tens of nanometers or even smaller for typical chiral ferromagnetic materials\cite{NatNano2013}. Since the period and decay length of the artifical microstructures are usually in the order of magnitude of micrometers\cite{AtomChip}, a much harder magnetic mirror for ultracold atoms can be realized with the chiral ferromagnetic film. Note that $\mathcal{B}_{h}(\mathbf{r})$ here is constant at given height $z$, which is different from the numerical result in Fig.~\ref{Fig2}(a). This minor deviation origins from the magnetic anisotropy energy in Eq.~(\ref{Ham}), where the single-$\mathbf{Q}$ state becomes an oversimplified and inaccurate description of the helical phase. It will be important to choose the materials with smaller magnetic anisotropy to get more smooth mirrors. 
     
The profile of magnetic field amplitude will be modified significantly when the bias magnetic field $\bm{\mathcal{B}}_{0}$ and the time-dependent magnetic field $\bm{\mathcal{B}}_{M}$ are turned on\cite{AtomChip,JPD1999}. Fig.~\ref{Fig2}(b) shows the distribution of total magnetic field amplitude $\mathcal{B}(\mathbf{r})$ for the device when $\bm{\mathcal{B}}_{0}=(0,0,20)$~mT and $\mathcal{B}_{M}=10$~mT. As shown in Fig.~\ref{Fig2}(b), the minima of $\mathcal{B}(\mathbf{r})$ in this case will locate at the parallel lines in the plane $z=8$~nm, and the appearing one-dimensional magnetic lattice is able to trap the ultracold atoms in the weak-field seeking state. We can define the ``recoil energy'' of the lattice as $E_{R}=\pi^{2}\hbar^{2}/(2m_{a}a^{2})$, where $m_{a}$ is the atom mass and $a$ is the lattice constant\cite{PRL2013}. For the helical phase with $a=42$~nm here, the recoil energy $E_{R}$ is estimated as $1.3$~neV for $^{87}\text{Rb}$ atom or $16$~neV for $^{7}\text{Li}$ atom respectively.

\begin{figure}
\includegraphics[scale=0.60]{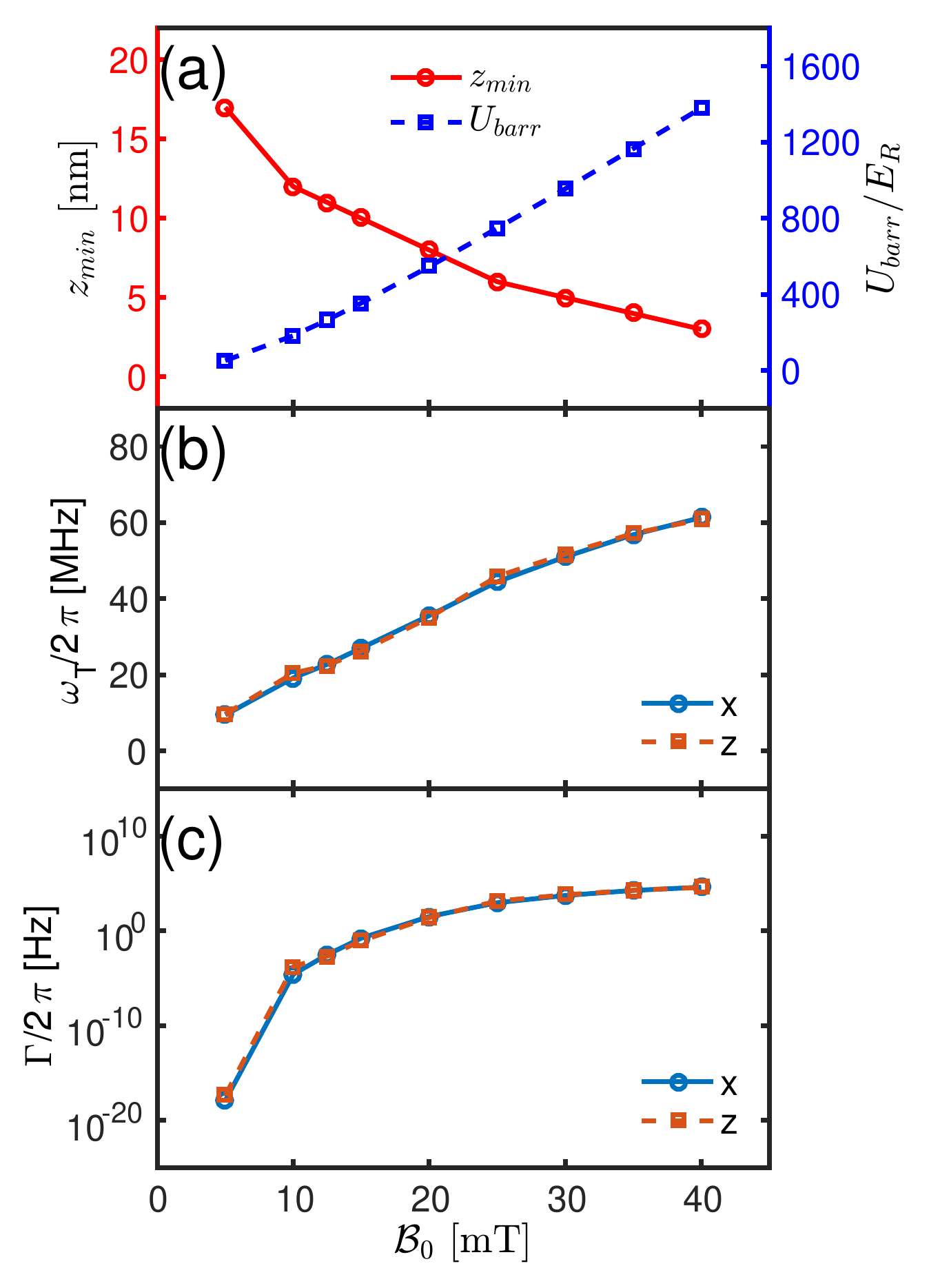}
\caption{(Color online) Field-dependent parameters of the one-dimensional magnetic lattice for $^{87}\text{Rb}$ atom. (a) trapping height $z_{\text{min}}$ and potential barrier $U_{\text{barr}}$ in the unit of the recoil energy $E_{R}$; (b) trapping frequency $\omega_{T}/2\pi$; (c) Marojana loss rate $\Gamma/2\pi$. Here, $\mathcal{B}_{0}$ varies from $5$~mT to $40$~mT, while $\mathcal{B}_{M}$ is fixed as $10$~mT. }\label{Fig3}
\end{figure}

The numerical results in Fig.~\ref{Fig2}(b) can also be understood with the help of the single-$\mathbf{Q}$ state. Assuming that $\bm{\mathcal{B}}_{0}=\frac{\mathcal{A}Q}{2}(0,0,B_{0})$ and $\mathcal{•}mathcal{B}_{M}=\frac{\mathcal{A}Q}{2}B_{M}$, the modulus of the total magnetic field $\bm{\mathcal{B}}(\mathbf{r})$ will be
\begin{eqnarray}
\mathcal{B}(\mathbf{r})=\frac{\mathcal{A}Q}{2}\sqrt{B_{0}^{2}+B_{M}^{2}+e^{-2Q|z|}+2B_{0}e^{-Q|z|}\cos(\mathbf{Q}\cdot\mathbf{r})}.\nonumber\\\label{grating}
\end{eqnarray} 
Therefore, such an operation will create an one-dimensional periodic potential for the ultracold atoms in the plane parallel to the film. This in fact forms an effective reflection grating for the incident atoms\cite{JPD1999}, which will attain the period in the range of nanometers and hence greatly extend the border of current technologies. Furthermore, the modulus $\mathcal{B}(\mathbf{r})$ will achieve its minimal value $\mathcal{B}_{\text{min}}=\mathcal{B}_{M}$ along the parallel lines defined by $\cos(\mathbf{Q}\cdot\mathbf{r}_{\text{min}})=-1,|z_{\text{min}}|=-\frac{1}{Q}\ln B_{0}$ in the case $B_{0}<1$. Thus, the magnetic field around each line will become a guide to trap the ultracold atoms, where the trapping height $|z_{\text{min}}|$ depends on the applied magnetic field. When $\mathcal{B}_{0}$ is continuously increased, the guides will get close to the film and finally disappear at $B_{0}=1$. This trend has also been verified in the numerical calculations, as shown in Fig.~\ref{Fig3}(a). When the magnetic field $\mathcal{B}_{0}$ varies from $5$~mT to $40$~mT, the trapping height of the one-dimensional magnetic lattice will decay from about $17$~nm to $3$~nm.   

In addition to the trapping height, the potential barrier for the atoms to escape from the minima locations $z_{\text{min}}$ to the infinitely far region $z\rightarrow\infty$ can also been tuned by the magnetic field $\bm{\mathcal{B}}_{0}$. The difference between the magnetic field amplitudes $\mathcal{B}(\infty)$ and $\mathcal{B}(z_{\text{min}})$ will be $\delta\mathcal{B}=\sqrt{\mathcal{B}_{0}^{2}+\mathcal{B}_{M}^{2}}-\mathcal{B}_{M}$, and the potential barrier will be $U_{\text{barr}}=m_{F}g_{F}\mu_{B}\delta\mathcal{B}$. When $\mathcal{B}_{M}$ is fixed as $10$~mT and $\mathcal{B}_{0}$ increases from $5$~mT to $40$~mT, the potential barrier for the $^{87}\text{Rb}$ and $^{7}\text{Li}$ atoms at $|F=2,m_{F}=2\rangle$ state will vary from $0.068$~$\mu$eV to $1.81$~$\mu$eV, which corresponds to the temperature range from $0.8$~mK to $21$~mK. In Fig.~\ref{Fig3}(a), the dependence of $U_{\text{barr}}$ on $\mathcal{B}_{0}$ is explicitly shown in the unit of the recoil energy $E_{R}$ for $^{87}\text{Rb}$ atom.

The trapping frequency $\omega_{T}$ of the atoms near the minima of the one-dimensional magentic lattice can be calculated as $\omega_{T}=\sqrt{m_{F}g_{F}\mu_{B}\mathcal{B}''_{\text{min}}/m_{a}}$, where $\mathcal{B}''_{\text{min}}$ is the magnetic field curvature at the minima. Fig.~\ref{Fig3}(b) shows the trapping frequencies along $x$ and $z$ directions for $^{87}\text{Rb}$ atom at $|F=2,m_{F}=2\rangle$ state when $\mathcal{B}_{M}=10$~mT and $\mathcal{B}_{0}\in [5,40]$~mT, where $\omega_{T,x}/2\pi\in[9.5,61.3]$~MHz and $\omega_{T,z}/2\pi\in[9.7,60.9]$~MHz respectively. Then the energy of the $^{87}\text{Rb}$ atom will be in the range $[0.04,0.25]$~$\mu$eV, which corresponds to the temperature range $[0.46,2.9]$~mT. Since the mass of $^{7}\text{Li}$ atom is about one-twenlfth of $^{87}\text{Rb}$ atom mass, its trapping frequency will be about $3.5$ times of the estimated value above.   

The Majorana loss rate near the minima of the lattice potential can be estimated as $\Gamma/2\pi=\omega_{T}\text{exp}(-4\omega_{L}/\omega_{T})$\cite{PRL2013,PRA2006}, where $\omega_{L}=m_{F}g_{F}\mu_{B}\mathcal{B}_{M}/\hbar$ is the Larmor precession frequency at the minima. As shown in Fig.~\ref{Fig3}(c), $\Gamma/2\pi$ for $^{87}\text{Rb}$ atom will increase from $10^{-18}$~Hz to $10^{4}$~Hz when the magnetic field $\mathcal{B}_{0}$ varies from $5$~mT to $40$~mT. Therefore, for given $\mathcal{B}_{M}$, the magnetic field $\mathcal{B}_{0}$ will give higher potential barrier $U_{\text{barr}}$ and larger Majorana loss rate $\Gamma/2\pi$ simultaneously, which should be optimally chosen during practical applications. Besides, the Majorana loss rate  for $^{7}\text{Li}$ atom will be larger because of its higher trapping frequency in the lattice.

More complex potential profiles for ultracold atoms can be constructed based on the helical phase in chiral ferromagnetic film. For example, if the applied magnetic field is along the magnetic stripes of the helical phase, the reflection plane of the magnetic mirror will be shifted toward the film\cite{JPD1999}. Then a vibrating magnetic mirror for the ultracold atoms can be realized if the applied field is harmonically oscillated\cite{JPD1999}. Furthermore, a ``moving grating" or ``conveyor belt"\cite{JPD1999} can be obtained if a magnetic field is rotating in the plane perpendicular to the magnetic stripes. Therefore, we expect that the helical phase in chiral ferromagnetic film can have widespread applications to develop and design various elementary devices for atom optics and atom chip.   

\section{Skyrmion Crystal Phase} 

\begin{figure*}
\includegraphics[scale=0.45]{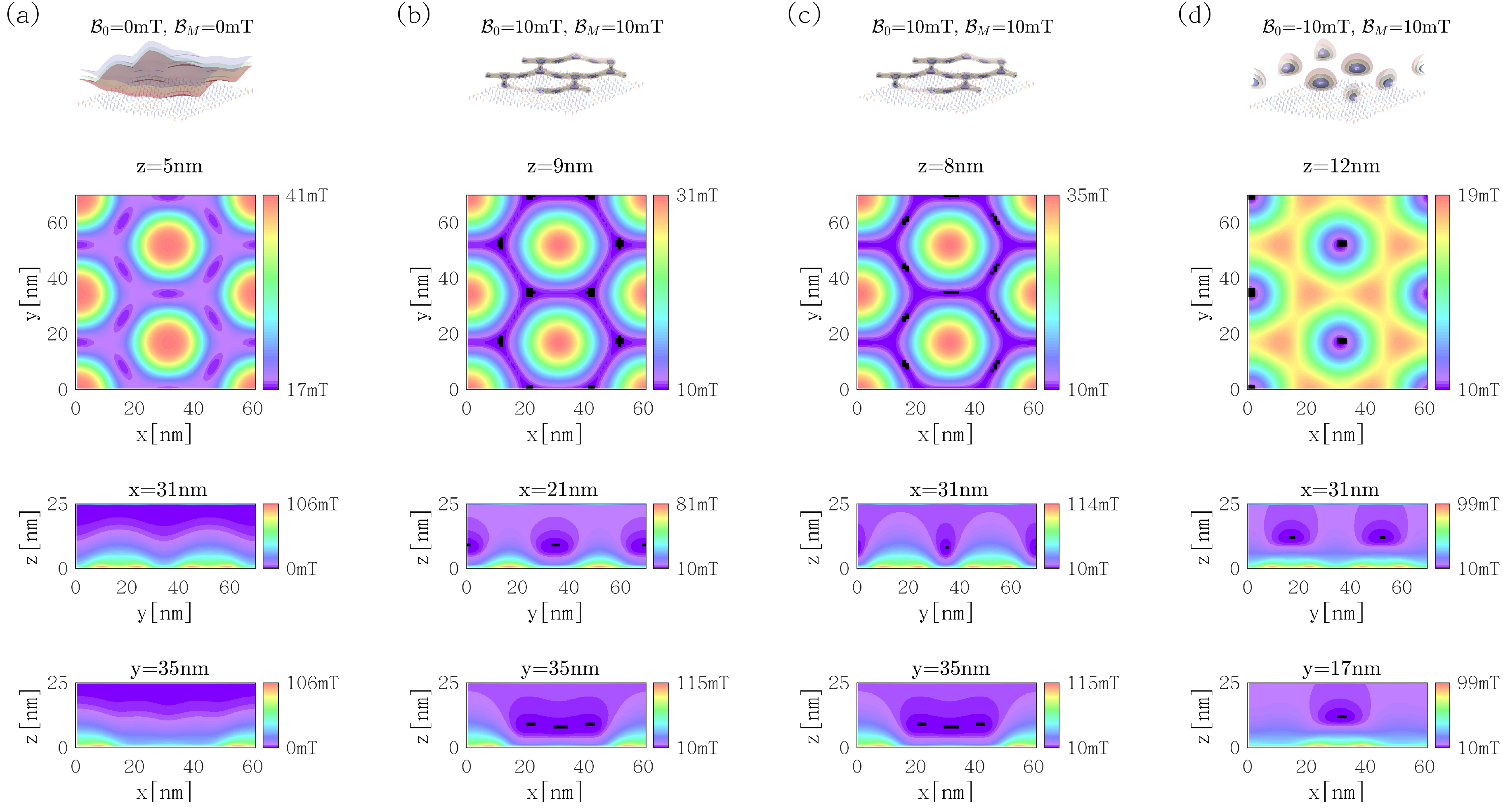}
\caption{(Color online) Profiles of magnetic field amplitude $\mathcal{B}(\mathbf{r})$ for the chiral ferromagnetic film at SkX phase. (a) Two-dimensional reflection gating when $\mathcal{B}_{0}=0$~mT and $\mathcal{B}_{M}=0$~mT. Top : isosurfaces of $\mathcal{B}(\mathbf{r})$ with three values $3$~mT, $6$~mT, and $9$~mT; middle and bottom : cross sections of $\mathcal{B}(\mathbf{r})$ when $z=5$~nm, $x=31$~nm, and $y=35$~nm respectively. (b) Honeycomb magnetic lattice when $\mathcal{B}_{0}=10$~mT and $\mathcal{B}_{M}=10$~mT. Top : isosurfaces of $\mathcal{B}(\mathbf{r})$ with three values $10.2$~mT, $10.4$~mT, $10.6$~mT; middle and bottom : cross sections of $\mathcal{B}(\mathbf{r})$ when $z=9$~nm, $x=21$~nm, and $y=35$~nm respectively. (c) Kagome magnetic lattice when $\mathcal{B}_{0}=10$~mT and $\mathcal{B}_{M}=10$~mT. Top : isosurfaces of $\mathcal{B}(\mathbf{r})$ with three values $10.2$~mT, $10.4$~mT, $10.6$~mT; middle and bottom : cross sections of $\mathcal{B}(\mathbf{r})$ when $z=8$~nm, $x=31$~nm, and $y=35$~nm respectively. (d) Triangular magnetic lattice when $\mathcal{B}_{0}=-10$~mT and $\mathcal{B}_{M}=10$~mT. Top : isosurfaces of $\mathcal{B}(\mathbf{r})$ with three values $11$~mT, $12$~mT, $13$~mT; middle and bottom : cross sections of $\mathcal{B}(\mathbf{r})$ when $z=12$~nm, $x=31$~nm, and $y=17$~nm respectively. }\label{Fig4}
\end{figure*}

The chiral ferromagnetic film can be driven into the SkX phase by a strong magnetic field $\mathbf{B}$, as shown in Fig.~\ref{Fig1}(b). Due to its topologically protected nature, the SkX phase will exist as a metastable state after withdrawing the applied magnetic field\cite{NatPhys2016}. The lifetime of the metastable SkX phase there can be as long as $10^{4}$~s at the temperature $23$~K\cite{NatPhys2016}, which is long enough to manipulate the ultracold atoms. The magnetic field distributions generated by Bloch-type and N\'{e}el-type SkXs have both been investigated thoroughly\cite{NJP2018}, which can be used to construct two-dimensional magnetic lattices for ultracold atoms. Here, we will focus on the Bloch-type SkXs given by Eq.~(\ref{Ham}), and the N\'{e}el-type SkXs can be treated in the same way.     

The magnetic field amplitude $\mathcal{B}_{c}(\mathbf{r})$ generated by the SkX phase in Fig.~\ref{Fig1}(b) without the external magnetic field $\mathbf{B}$ is shown in Fig.~\ref{Fig4}(a). Here, $\mathcal{B}_{c}(\mathbf{r})$ will decrease from $105$~mT to $1$~mT when the height $z$ increases from $0$~nm to $25$~nm; while in the plane at height $z=5$~nm, $\mathcal{B}_{c}(\mathbf{r})$ will periodically varies from $18$~mT to $40$~mT. Specially, the minima of $\mathcal{B}_{c}(\mathbf{r})$ at fixed height form a Kagome lattice, while its maxima at the same height form a triangular lattice. Therefore, the chiral ferromagnet film in the SkX phase can be viewed as a two-dimensional ``reflection grating" for the ultracold atoms. 

When the bias magnetic field $\bm{\mathcal{B}}_{0}$ and the time-dependent magnetic field $\bm{\mathcal{B}}_{M}$ are switched on, the minima of the total magnetic field amplitude $\mathcal{B}(\mathbf{r})$ will appear at finite height, which will form two-dimensional magnetic lattices to trap the ultracold atoms. For example, the distribution of $\mathcal{B}(\mathbf{r})$ when $\mathcal{B}_{0}=\mathcal{B}_{M}=10$~mT is shown in Fig.~\ref{Fig4}(b) and (c). Here, one group of minimal points appear at the height $z_{\text{min}}=9$~nm and locate at the centers of the triangles determined by any three nearest skyrmions, which thus form a honeycomb magnetic lattice. Another group of minimal points locate at the middle points between any two nearest skyrmions at the height $z_{\text{min}}=8$~nm, which thus form a Kagome magnetic lattice. When the direction of the bias magnetic field is reversed with $\mathcal{B}_{0}=-10$~mT, a triangular magnetic lattice will be formed at the height $z_{\text{min}}=12$~nm, and the minimal points of $\mathcal{B}(\mathbf{r})$ will locate at the center of the skyrmions, as shown Fig.~\ref{Fig4}(d). The recoil energy $E_{R}$ now will be $1.9$~neV for $^{87}\text{Rb}$ atom and $24$~neV for $^{7}\text{Li}$ atom, where the lattice constant for SkX phase is about $35$~nm.
     
The magnetization configuration of the Bloch-type SkX is approximately described by a triple-$\mathbf{Q}$ state\cite{NatNano2013,NJP2018}
$\mathbf{m}_{skx}(\mathbf{r})=\mathbf{m}_{0}(\mathbf{r})\delta(z)
+\mathcal{A}\sum\limits_{i=1}^{3}[\hat{\mathbf{e}}_{z}\cos(\mathbf{Q}_{i}\cdot\mathbf{r})+\hat{\mathbf{e}}_{i}\sin(\mathbf{Q}_{i}\cdot\mathbf{r})]\delta(z)$. Without loss of generality, the wavevectors $\mathbf{Q}_{i}=Q\hat{\mathbf{Q}}_{i}$ here are set as $\hat{\mathbf{Q}}_{1}=(1,0,0),\hat{\mathbf{Q}}_{2}=(-\frac{1}{2},\frac{\sqrt{3}}{2},0),\hat{\mathbf{Q}}_{3}=(-\frac{1}{2},-\frac{\sqrt{3}}{2},0)$, which then give the set of unit vectors $\hat{\mathbf{e}}_{i}=\hat{\mathbf{e}}_{z}\times\hat{\mathbf{Q}}_{i}$. The spatial distribution of magnetic field generated by Bloch-type SkX is explicitly expressed as\cite{NJP2018}
\begin{eqnarray}
&&\bm{\mathcal{B}}_{skx}(\mathbf{r})=\frac{\mathcal{A}Q}{2}e^{-Q|z|}\nonumber\\&&\times\left(\begin{array}{c}\text{sgn}(z)(\sin(Qx)+\sin(\frac{1}{2}Qx)\cos(\frac{\sqrt{3}}{2}Qy))\\\text{sgn}(z)\sqrt{3}\cos(\frac{1}{2}Qx)\sin(\frac{\sqrt{3}}{2}Qy)\\\cos(Qx)+2\cos(\frac{1}{2}Qx)\cos(\frac{\sqrt{3}}{2}Qy)
\end{array}\right).\label{Bskx}
\end{eqnarray}
Its modulus is 
$\mathcal{B}_{skx}(\mathbf{r})=\frac{\mathcal{A}Q}{2}e^{-Q|z|}\sqrt{\mathcal{F}(x,y)}$, where $\mathcal{F}(x,y)=1+3\cos^{2}(\frac{1}{2}Qx)+\cos^{2}(\frac{\sqrt{3}}{2}Qy)+4\cos^{3}(\frac{1}{2}Qx)\cos(\frac{\sqrt{3}}{2}Qy)$. We see that $\mathcal{B}_{skx}(\mathbf{r})$ will decay exponentially along with the height $|z|$. Besides, $\mathcal{F}(x,y)$ will reach it minimal value $1$ on a Kagome lattice and reach its maximal value $9$ on a triangular lattice. Therefore, the analytical results based on the triple-$\mathbf{Q}$ state here coincide with the features revealed numerically in Fig.~\ref{Fig4}(a).
  
\begin{figure}
\includegraphics[scale=0.60]{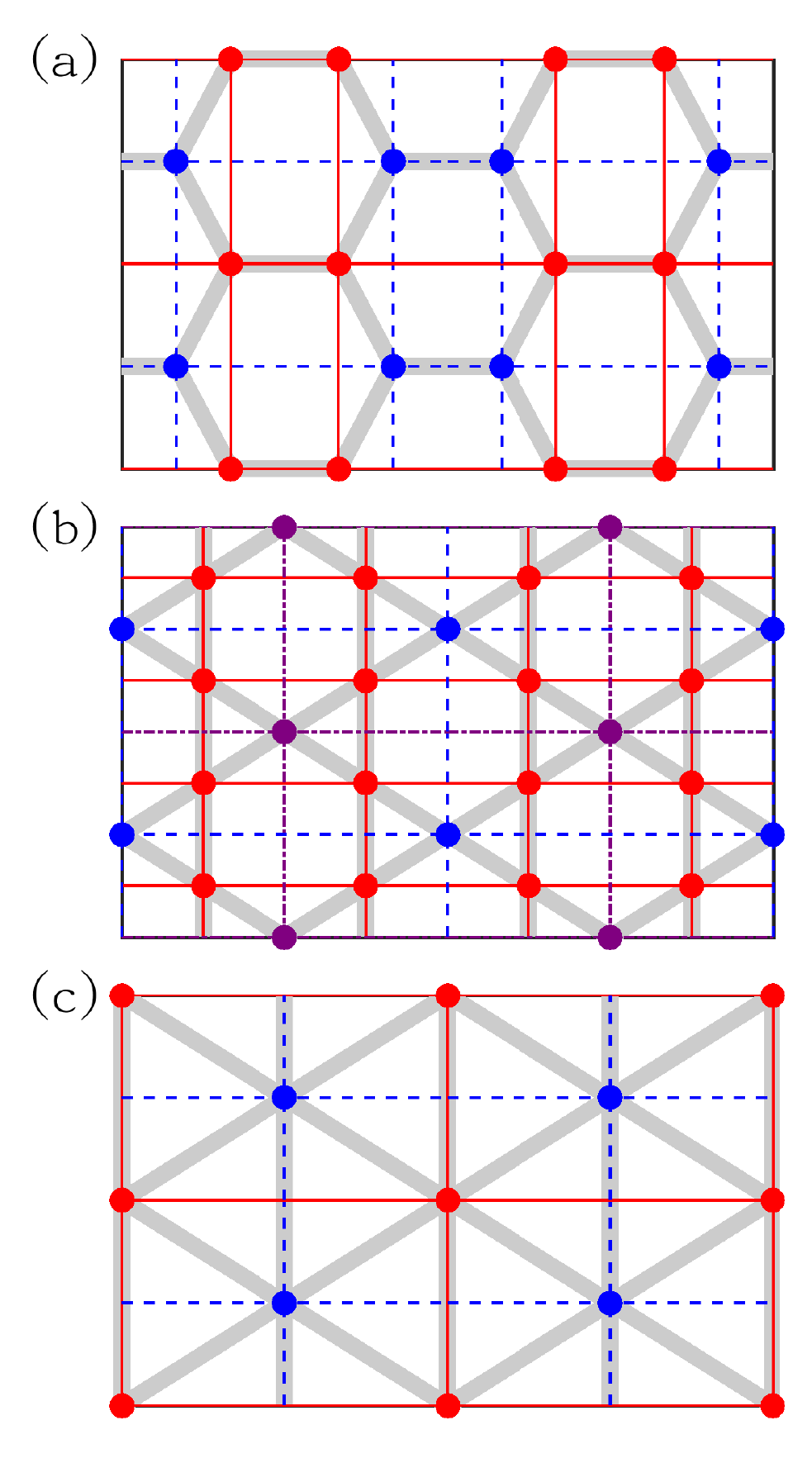}
\caption{(Color online) The sites of three types of two-dimensional magnetic lattices, which are determined by the three sets of solutions of Eq.~(\ref{Beq0}). (a) Honeycomb lattice; (b) Kagome lattice; (c) triangular lattice.}\label{Fig5}
\end{figure}  
  
After turning on the magnetic fields $\mathcal{B}_{0}=\frac{\mathcal{A}Q}{2}B_{0}$ and $\mathcal{B}_{M}=\frac{\mathcal{A}Q}{2}B_{M}$, which are weak enough that the SkX phase is not destroyed, the modulus of the total magnetic field will be $\mathcal{B}(\mathbf{r})=\frac{\mathcal{A}Q}{2}\sqrt{\mathcal{G}(\mathbf{r})}$, where $\mathcal{G}(\mathbf{r})=B_{0}^{2}+B_{M}^{2}+2B_{0}e^{-Q|z|}[\cos(Qx)+2\cos(\frac{1}{2}Qx)\cos(\frac{\sqrt{3}}{2}Qy)]+e^{-2Q|z|}\mathcal{F}(x,y)$. Hence, $\mathcal{B}(\mathbf{r})$ will be finite and homogeneously distributed as $|z|\rightarrow\infty$, while the minimal value $\mathcal{B}_{\text{min}}=10$~mT of $\mathcal{B}(\mathbf{r})$ can be achieved at periodic arrays of points at finite height $z_{\text{min}}$. The minimal points of $\mathcal{B}(\mathbf{r})$ will be determined by 
\begin{eqnarray}
\left\{
\begin{aligned}
\sin(\frac{1}{2}Qx)(2\cos(\frac{1}{2}Qx)+\cos(\frac{\sqrt{3}}{2}Qy))&=&0,\\
\cos(\frac{1}{2}Qx)\sin(\frac{\sqrt{3}}{2}Qy)&=&0,\\
e^{-Q|z|}(\cos^{2}(\frac{1}{2}Qx)-\sin^{2}(\frac{1}{2}Qx)\quad\quad&&\\
+2\cos(\frac{1}{2}Qx)\cos(\frac{\sqrt{3}}{2}Qy))+B_{0}&=&0.
\end{aligned}\right.\label{Beq0}
\end{eqnarray}  

\begin{figure*}
\includegraphics[scale=0.60]{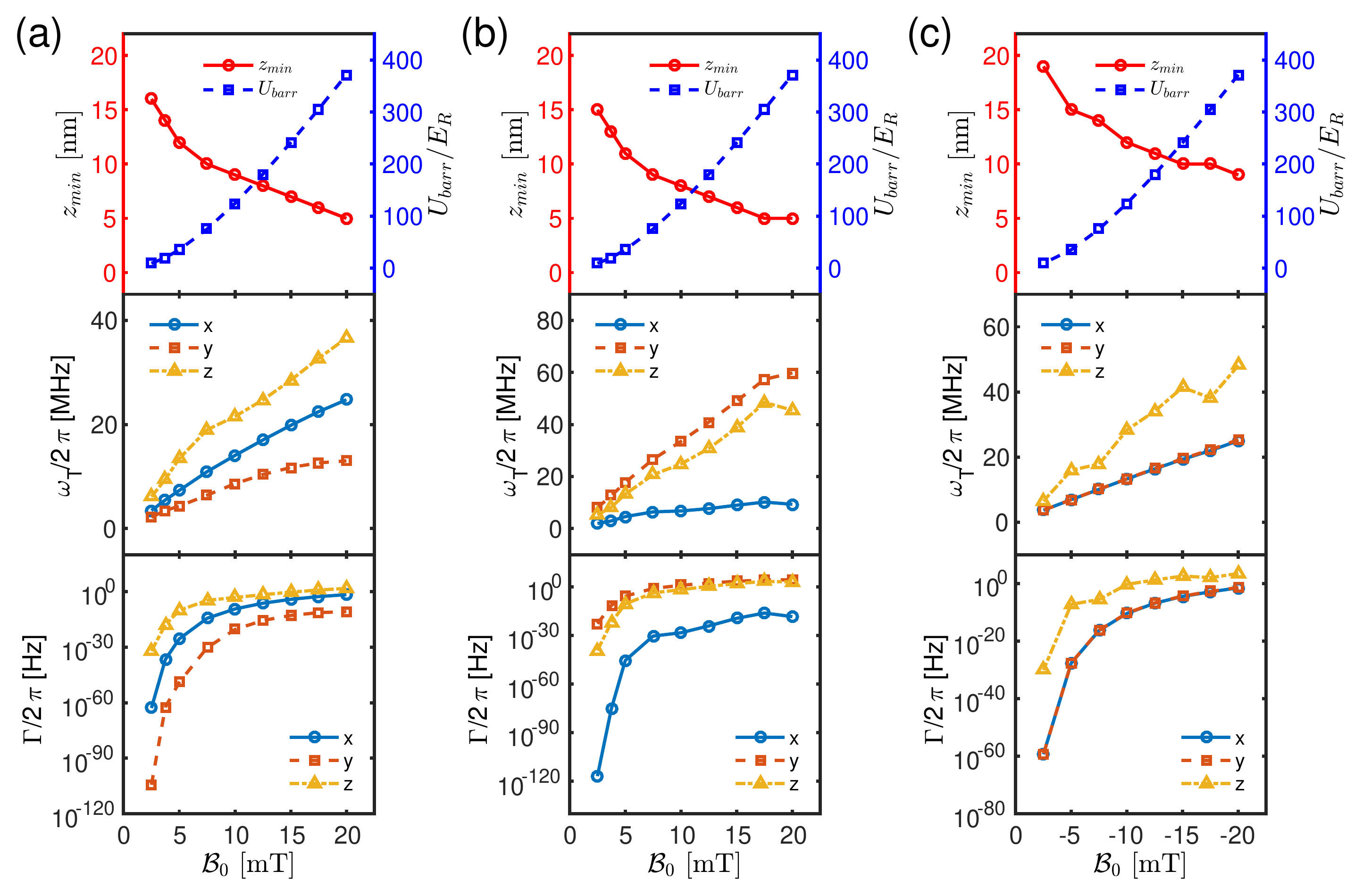}
\caption{(Color online) The dependence of trapping height $z_{\text{min}}$, potential barrier $U_{\text{barr}}$, trapping frequency $\omega_{T}/2\pi$, and Marojana loss rate $\Gamma/2\pi$ for $^{87}\text{Rb}$ atom on the bias magnetic field. Here, $E_{R}$ is the recoil energy of $^{87}\text{Rb}$ atom, $\mathcal{B}_{0}$ varies from $2.5$~mT to $20$~mT, and $\mathcal{B}_{M}$ is fixed as $10$~mT. (a) Honeycomb lattice; (b) Kagome lattice; (c) triangular lattice.}\label{Fig6}
\end{figure*} 

Three sets of solutions are found for Eq.~(\ref{Beq0}), which then define three types of two-dimensional magnetic lattices for the ultracold atoms:  

(I) Honeycomb lattice if $0<B_{0}<3/2$. The lattice sites are defined by $\sin(\frac{1}{2}Qx)=\pm\frac{\sqrt{3}}{2},\quad\cos(\frac{1}{2}Qx)=-\frac{1}{2},\quad\sin(\frac{\sqrt{3}}{2}Qy)=0,\quad\cos(\frac{\sqrt{3}}{2}Qy)=1,\quad e^{-Q|z|}=\frac{2}{3}\mathcal{B}_{0}$, or $\sin(\frac{1}{2}Qx)=\pm\frac{\sqrt{3}}{2},\quad\cos(\frac{1}{2}Qx)=\frac{1}{2},\quad\sin(\frac{\sqrt{3}}{2}Qy)=0,\quad\cos(\frac{\sqrt{3}}{2}Qy)=-1,\quad e^{-Q|z|}=\frac{2}{3}B_{0}$, as shown in Fig.~\ref{Fig5}(a). The height of these points is $|z_{\text{min}}|=-\frac{1}{Q}\ln(\frac{2}{3}B_{0})$, which will decrease with increased $B_{0}$ and vanish at $B_{0}=3/2$. The honeycomb optical lattices have been proposed\cite{PRA2009} and realized to trap the ultracold atoms, which then are able to simulate the exotic phenomena including the superfluid-to-Mott-insulator transition\cite{NatPhys2011-2}, Dirac points\cite{Nature2012}, quantum anomalous Hall effect\cite{Nature2014}, \emph{etc}. Therefore, similar physical phenomena of the ultracold atoms are expected when they are trapped by the honeycomb magnetic lattice here.  

(II) Kagome lattice if $0<B_{0}<1$. The lattice sites are defined by 
$\sin(\frac{1}{2}Qx)=\pm 1,\quad\cos(\frac{1}{2}Qx)=0,\quad\sin(\frac{\sqrt{3}}{2}Qy)=\pm 1\quad\cos(\frac{\sqrt{3}}{2}Qy)=0,\quad e^{-Q|z|}=B_{0}$, or
$\sin(\frac{1}{2}Qx)=0,\quad\cos(\frac{1}{2}Qx)=1,\quad\sin(\frac{\sqrt{3}}{2}Qy)=0,\quad\cos(\frac{\sqrt{3}}{2}Qy)=-1,\quad e^{-Q|z|}=B_{0}$, or
$\sin(\frac{1}{2}Qx)=0,\quad\cos(\frac{1}{2}Qx)=-1,\quad\sin(\frac{\sqrt{3}}{2}Qy)=0,\quad\cos(\frac{\sqrt{3}}{2}Qy)=1,\quad e^{-Q|z|}=B_{0}$, as shown in Fig.~\ref{Fig5}(b). The height of these points is $|z_{\text{min}}|=-\frac{1}{Q}\ln B_{0}$, which will also decrease with increased $B_{0}$ and vanish at $B_{0}=1$. Therefore, the Kagome lattice is more close to the film surface than the honeycomb lattice when the two coexist, which is also revealed in Fig.~\ref{Fig4}(b)(c). Kagome lattice has the intriguing ability to host flat band states of ultracold atoms, which is an ideal platform to study quantum many-body physics. Hence, the chiral ferromagnetic film provides a novel platform to realize the Kagome lattice for ultracold atoms in addition to the optical method\cite{PRL2009,PRL2012}.   

(III) Triangular lattice if $-3<B_{0}<0$. The lattice sites are defined by $
\sin(\frac{1}{2}Qx)=0,\quad\cos(\frac{1}{2}Qx)=1,\quad\sin(\frac{\sqrt{3}}{2}Qy)=0,\quad\cos(\frac{\sqrt{3}}{2}Qy)=1,\quad e^{-Q|z|}=-\frac{1}{3}B_{0}$,
or $\sin(\frac{1}{2}Qx)=0,\quad\cos(\frac{1}{2}Qx)=-1,\quad\sin(\frac{\sqrt{3}}{2}Qy)=0,\quad\cos(\frac{\sqrt{3}}{2}Qy)=-1,\quad e^{-Q|z|}=-\frac{1}{3}B_{0}$, as shown in Fig.~\ref{Fig5}(c). The height of these points is $|z_{\text{min}}|=-\frac{1}{Q}\ln(-\frac{1}{3}B_{z})$, which will decrease with increased $B_{0}$ and vanish at $B_{0}=-3$. Similar to the case of optical lattice, the triangular magnetic lattice here can also be useful to simulate geometricaly frustrated magnetism with ultracold atoms\cite{Science2011}.   

The properties of the two-dimensional magnetic lattices are further investigated for the chiral ferromagnetic film described by Eq.~(\ref{Ham}). As shown in Fig.~\ref{Fig6}, the trapping height of each lattice will decrease when the bias magnetic field $\mathcal{B}_{0}$ is continuously increased with fixed $\mathcal{B}_{M}=10$~mT, which agree with the analytical results above. Meanwhile, the potential barrier $U_{\text{barr}}$ for the same type of atom will be the same as the one-dimensional magnetic lattice discussed in Section III, since $\delta\mathcal{B}$ only depends on $\mathcal{B}_{0}$ and $\mathcal{B}_{M}$. The trapping height and potential barrier will be the same for $^{7}\text{Li}$ atom at the state $|F=2,m_{F}=2\rangle$.   

The ultracold atoms will be confined in three directions by the two-dimensional magnetic lattices, and the trapping frequencies $\omega_{T}$ for $^{87}\text{Rb}$ atoms are given in Fig.~\ref{Fig6} when $\mathcal{B}_{M}=10$~mT and $|\mathcal{B}_{0}|$ varies from $2.5$~mT to $20$~mT. Similar to the one-dimensional magnetic lattice, stronger bias magnetic field will induce higher trapping frequency here. Besides, the inequivalence of trapping frequencies along different direction suggest the anisotropy of the trapping potential near the minima of the two-dimensional magnetic lattices. The Majorana loss rates $\Gamma/2\pi$ for $^{87}\text{Rb}$ atom in the two-dimensional magnetic lattices are also presented in Fig.~\ref{Fig6}. As expected, $\Gamma/2\pi$ will be enhanced with increasing bias magnetic field $\mathcal{B}_{0}$, and has the same anisotropic feature as the trapping frequency $\omega_{T}$. Similar to the one-dimensional magnetic lattice, $^{7}\text{Li}$ atom at the state $|F=2,m_{F}=2\rangle$ will attain higher trapping frequency and larger Majorana loss rate due to its smaller mass.
 
\section{Discussions and Conclusion} 
Although the chiral ferromagnetic film has shown attractive features to control the ultracold atoms, there are still several issues to be addressed for practical performances. First of all, one needs to transfer the ultracold atoms to the magnetic lattices near the film surface for further operations. This might be accomplished by loading the atoms from an external dimple trap\cite{DimpleTrap} by adiabatically tuning the trapping height with the bias field $\mathcal{B}_{0}$, which in fact has been proposed for the magnetic lattice created by a superconducting film\cite{PRL2013}. Secondly, imperfections will always be introduced to the magnetic field by the unavoidable defects in chiral ferromagnetic film, such as the bent helical stripes, irregular array of skyrmions, or even their mixtures\cite{NatNano2013}. In order to get ideal magnetic lattices, it will be critical to improve the quality of chiral ferromagnetic film and get perfect magnetic patterns. Finally, the trapping height of the ultracold atoms will be tens of nanometers, which is so close that the Casimir-Polder force between the atoms and the film surface will be significantly enhanced\cite{AtomChip}. Therefore, the dynamic behavior of ultracold atoms interacting with the chiral ferromagnetic film should be highly different from the case of optical lattice, which needs to be explored further.   

In conclusion, we have investigated the possibility to control ultracold atoms with the magnetic field produced by the chiral ferromagnetic film. We demonstrate how to realize magnetic mirror, reflection grating, one-dimensional and two-dimensional magnetic lattices in the proposed device. Compared with current top-down techniques to produce magnetic fields for ultracold atoms, the strategies based on chiral ferromagnetic film here belong to the bottom-up category and will be more flexible and controllable. With the benefit of the continuous advances of the spintronics technology and material science, the interactive ultracold atoms and chiral ferromagnetic film can be a promising platform to demonstrate exotic physics phenomena and develop novel quantum techniques.

\section*{Acknowledgements}                  
This work has been supported by NSFC Project No. 61674083 and No. 11604162, and ``the Fundamental Research Funds for the Central Universitie'' Nankai University (7540).

\end{document}